# How Good Can 2D Excitonic Solar Cells Be?

Zekun Hu,[1] Da Lin,[2] Jason Lynch,[1] Kevin Xu,[1] Deep Jariwala[1,*]

[1]*Department of Electrical and Systems Engineering, University of Pennsylvania, Philadelphia, PA, 19104, USA*

[2]*Department of Materials Science and Engineering, University of Pennsylvania, Philadelphia, PA, 19104, USA*

*Corresponding Author (D.J.) email: dmj@seas.upenn.edu

## Abstract

**Excitonic semiconductors have been a subject of research for photovoltaic applications for many decades. Among them, the organic polymers and small molecules based solar cells have now exceeded 19% power conversion efficiency (PCE). While organic photovoltaics (OPVs) are approaching maturity, the advent of strongly excitonic inorganic semiconductors such as two-dimensional transition metal dichalcogenides (TMDCs) has renewed interest in excitonic solar cells due to their high-optical constants, stable inorganic structure and sub-nm film thicknesses. While several reports have been published on TMDC based PVs, achieving power conversion efficiencies higher than 6% under one-sun AM1.5G illumination has remained challenging. Here, we perform a full optical and electronic analysis of design, structure and performance of monolayer TMDC based, single-junction excitonic PVs. Our computational model with optimized properties predicts a PCE of 9.22% in a superlattice device structure. Our analysis suggests that, while the PCE for 2D excitonic solar cells may be limited to < 10%, a specific power > 100 W g$^{-1}$ may be achieved with our proposed designs, making them attractive in aerospace, distributed remote sensing, and wearable electronics.**

## Introduction

Thin-film photovoltaics, such as those based on III-V semiconductors, CdTe, and 3D perovskites, have been a source of sustained research and commercial interest. However, they occupy a small share of the large-scale, grid-tied market since their production has not been scaled. Hence, their price of electricity remains high compared to silicon photovoltaics (PVs) which is the dominant PV technology[1]. However, thin-film photovoltaics have long been considered as a potential solution for lightweight applications, such as aerospace, powering distributed remote sensors, and wearable electronics[2, 3]. In this application, new and emerging materials such as organic semiconductors[4], II-chalcogenides[5] and two-dimensional hybrid organic-inorganic perovskites[6] are also being heavily investigated. Among novel, thin-film photovoltaic materials, excitonic semiconductors have attracted a lot of attention due to their large absorption coefficients which permit a sharp reduction in active layer thickness of the PV devices. However, 2D transition metal dichalcogenides (TMDCs) of $MX_2$ (M=Mo, W and X=S, Se, Te) have recently gained traction for lightweight PV applications. In particular, their large optical constants result in large loss-tangent values across the visible region combined with their availability in high optical and electronic quality over wafer scales makes then increasingly viable candidates for thin-film, ultralight-weight PVs[7, 8].

Further, a notable feature of 2D TMDC semiconductors is a transition from an indirect bandgap in bulk to the direct bandgap in monolayers that enables a high photoluminescence quantum yield, and thus, a high radiative efficiency[9]. Finally, the wide range of bandgaps (1.0-2.5 eV[10]) and van der Waals bonding for facile hetero-integration make TMDCs attractive candidates for single-junction, tandem, and multi-junction solar cells[11].

Consequently, several studies reporting microscale PV devices from bulk and monolayer TMDCs are available. However, the PCE values of these experimentally reported TMDC solar cells are typically lower than 2%[12-14], whereas the highest PCEs were reported as 9.03% in $MoS_2$[15] and 6.3% in $WS_2$[16]. Per the detailed balance model[11], thin-film, single-junction TMDC solar cells can have maximum PCEs of up to 27%, comparable to crystalline Silicon. This vast disparity between theoretical maximum and experimentally observed values therefore merits further investigation. To investigate the practically limiting parameters of current TMDC PVs, we developed a combined optical and electronic model to simulate the photovoltaic characteristics of four TMDC materials ($MoS_2$, $WS_2$, $MoSe_2$, and $WSe_2$) based on an experimentally reported, large area, scalable superlattice structure[17]. Based on our model, we optimized the parameters and performance of monolayer $MoS_2$ superlattice based PV devices, attaining a PCE as high as 9.22% under one-sun (AM1.5) illumination with specific power exceeding 50 W/g. Finally, we benchmarked our results against other 2D PV devices on the metrics of PCE vs specific power and find that 2D TMDC based PVs when optimized for both optical and electronic design can outperform all available technologies in high specific power applications.

# Results and Discussion

**Device Structure and Optimization of Photon Absorption:**

The proposed 2D TMDC-based photovoltaic superlattice device is shown in Figure 1a. The device consists of a repeating unit cell of a monolayer $MoS_2$ absorber (0.65 nm) and an $Al_2O_3$ insulator (3nm), and is placed on top of an $Al_2O_3$/Au substrate with the Au serving as a reflector. The thickness of the $Al_2O_3$ layer has been optimized to enhance photocarrier generation. The active layer of the device is intrinsic (no doping) which is 1 μm long, with silver and gold cathode and anode electrodes measuring 0.01 μm in length each. The heavily doped p-region ($10^{19}$ $cm^{-3}$) near the cathode and the heavily doped n-region ($10^{19}$ $cm^{-3}$) near the anode are also 0.01 μm in length. The incident angle of the light is zero, meaning it is normal to the surface. Additional details regarding the device structure and parameters used for simulation are provided in methods and in supporting information Table S1.

By using the Transfer Matrix Method (TMM) to calculate the absorbed photon density of each layer, we find that increasing the number of layers significantly increases the absorbed photon density in the 400-700 nm wavelength range as seen in Figure 1c. This increase is particularly significant when going from N=1 to N=5. However, further increasing the number of layers from N=5 to N=10 does not lead to as dramatic an increase in broadband absorptance and actually decreases the absorbed energy per unit weight. Therefore, the electrical simulations for our proposed device were based on an N=5 superlattice design. The energy band diagram for the active layer, which has a 1.80 eV bandgap for monolayer $MoS_2$[18], is shown in Figure 1d. The band diagram is that of a p-i-n lateral homojunction takes into account the built-in field created by degenerate p-doping and n-doping near the contacts which promotes charge carrier selectivity and improves device performance[19].

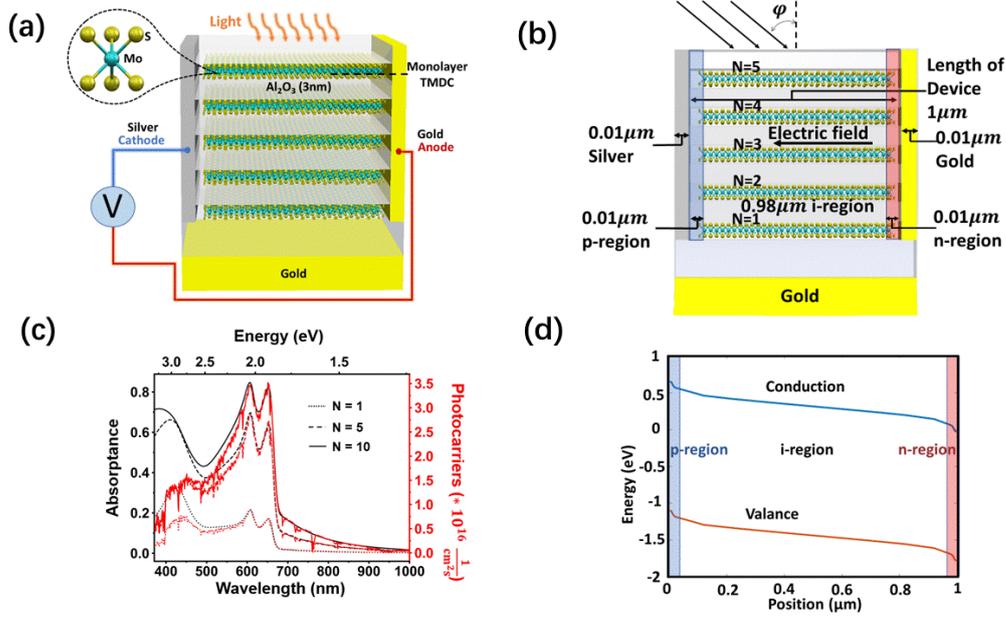

Figure 1. (a) Schematic model of the superlattice structure. (b) Side view of the model with labeled p-i-n regions and layers. (c) Simulated absorptance spectra and absorbed photon density for $MoS_2/Al_2O_3$ (N=1,5,10). (d) Energy band diagram of the monolayer $MoS_2$ for p-i-n junction.

**Role of Excitons and their Radiative Efficiency**

It is well known that excitons dominate the optical response of semiconducting 2D TMDCs, not only in the monolayer limit but also in the bulk. However, thus far, all attempts in literature to quantify and estimate theoretical PV potential for TMDCs have failed to treat excitons seriously in their models[20-22]. Not only do excitons dominate the optical properties in low dimensional semiconductors but it is also well-known through both theory and experiment that they limit the performance of other well-known excitonic semiconductors-based PVs, namely OPVs[23-28]. Therefore, any attempt to seriously quantify the performance limits of semiconducting 2D TMDC based PVs must include a detailed treatment of excitons. In this work, we have thus examined the effects of exciton binding energy (BE), exciton radiative lifetime ($\tau_{ex-r}$), exciton nonradiative lifetime ($\tau_{ex-nr}$), and exciton diffusion length (EDL) in our model for determining performance limits of semiconducting 2D TMDC based PVs. The binding energy of excitons, which describes the ease with which an electron and hole can be separated, is important for solar cell performance. In $MoS_2$, the exciton binding energy increases from 0.08 eV[29] in the bulk to 0.44 eV[30] in monolayers due to quantum confinement effects[31]. However, the binding energy in monolayer TMDCs can be modified by molecular coverage[32], doping[33], and engineering of the dielectric environment[34]. By simulating the PCE as a function of exciton binding energy and diffusion length (Figure 2a and Table S2), we found that higher PCE was observed at lower binding energies. This is because excitons are more likely to dissociate at the interface with the aid of thermal energy as the binding energy decreases (Figure S1a,b).

EDL is another crucial parameter that influences the transportation of excitons in monolayer TMDCs and therefore affects solar cell performance. Since excitons are neutral, they are not significantly affected by electric field drift. In organic solar cells, the diffusion length is on the nanometer scale, making it unlikely for an exciton to reach an electrode[35]. In contrast, the diffusion length of TMDCs is in the micron range, allowing the electrodes to be further apart and reducing the amount of reflected light. The simulated range of 0.015-6 μm encompasses previously measured diffusion lengths in $MoS_2$[36]. Assuming that excitons follow a Gaussian distribution centered around the diffusion length and are excited evenly throughout the $MoS_2$, photocurrent is still produced even when the diffusion length is less than the distance between electrodes (1 μm). However, some excitons will still recombine before being converted into photocurrent. In silicon PVs, the recombination process releases heat due to the indirect band gap, but in our solar cell design with a direct band gap, a photon is emitted during recombination instead of heat, which does not reduce overall efficiency. The emitted photon may either be radiated out of the solar cell or reabsorbed. We did not consider the effects of reabsorption[37] due to our expectation that they would be low. In the blue curve of Figure 2b (0.4 eV binding energy), the PCE increases with diffusion length from 0.015 μm to 2 μm as a longer diffusion length increases the probability of an exciton reaching one of the interfaces of the junction. However, when the diffusion length becomes larger than twice the device length, nearly all the excitons reach the interfaces and begin to concentrate at the electrodes, blocking other excitons from disassociating at the interface and decreasing the PCE. For large binding energies (BE > 0.3 eV), the results for PCE, short-circuit current ($I_{SC}$), and open-circuit voltage ($V_{OC}$) were similar to the blue curve as shown in Figure S1c-e. For the red curve with a relatively low binding energy of 0.2 eV, the higher number of dissociated charges almost fully occupy the space between electrodes, resulting in a decrease in PCE as the diffusion length increases. Further increase of diffusion length results in a purely declining trend of PCE for all binding energies < 0.3 eV.

To understand the factors influencing exciton diffusion, we examined the radiative lifetime and nonradiative lifetime of the excitons based on a binding energy of 0.24 eV[38] and a diffusion length of 1.5 μm. The quantum yield of monolayer $MoS_2$ can range from less than 1% to nearly 100%, and the effective lifetime can be as long as 10 ns[39]. We first conducted simulations varying the exciton radiative lifetime from 0.0001-10 ns and the exciton nonradiative lifetime from 0.0001-10 ns, which modified the quantum yield from less than 1% to greater than 99% (Figure 2c and Figure S2a,b). Previous reports have also found radiative lifetimes ranging from 0.2 ps[40] to 15 ns[41] and nonradiative lifetimes from 1 ps to 10 ps[42] in $MoS_2$. In order to more closely examine the effect of these lifetimes on PCE, we further studied the range with logarithmic axes in Figure 2d and Figure S2c,d. As the simulations did not take into account the reabsorption of photons from recombination, the quantum yield had no impact on PCE.

The total lifetime, which is the combination of radiative and nonradiative lifetimes, determines the photocurrent and PCE. A longer lifetime means excitons exist longer before recombining and have a higher chance of diffusing and dissociating, leading to higher photocurrent from an increase in charges. In Figure 2d, improving the nonradiative lifetime is more important than improving the radiative lifetime as defects significantly trap the excitons and reduce the total lifetime, and therefore the efficiency of the photovoltaic device. In the optimization process, we selected 10 ns for both radiative and nonradiative lifetimes, resulting in a PCE of 5.62%.

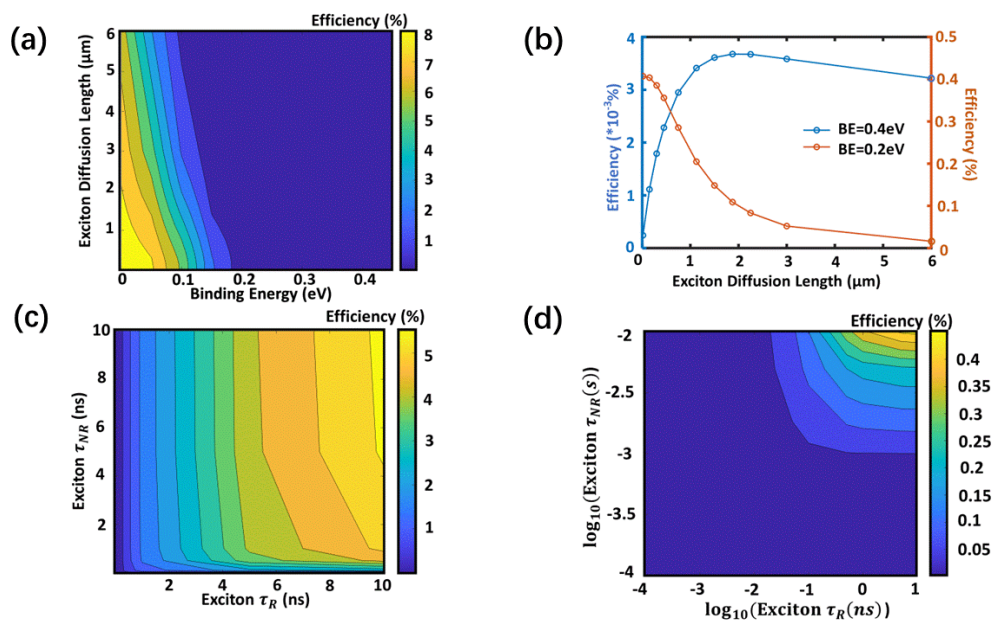

Figure 2. (a) The PCE for the variation of exciton binding energy (0-0.5 eV) and diffusion length (0.015-6 μm). (b) The PCE for the variation of exciton diffusion length for BE=0.2 eV, 0.4 eV. (c) The PCE for the variation of eτ$_{ex-r}$ and τ$_{ex-nr}$ (0.0001-10 ns). (d) The current density for the variation of eτ$_{ex-r}$ (0.0001-0.01 ns) and τ$_{ex-nr}$ (0.0001-10 ns).

**Role of free carriers and their mobilities**

After optimizing the parameters of exciton binding energy and exciton lifetimes, we examined the effect of free carrier properties under an electric field on photocurrent. The electron mobility of monolayer MoS$_2$ has been reported to be in the range of 0.1-10 cm$^2$V$^{-1}$s$^{-1}$ [43, 44]. However, the carrier mobility can be improved using the dielectric screening effect[45], allowing for values greater than 100 cm$^2$V$^{-1}$s$^{-1}$ [46, 47]. We therefore studied the range of free carrier mobilities from 0.1 to 200 cm$^2$V$^{-1}$s$^{-1}$ and device lengths from 0.4 to 10 μm (Figure 3a and Figure S3). As the electron mobility increases, the PCE is enhanced as more photocarriers are collected by the electrodes. In Figure 3b, we see that the PCE increases significantly in the mobility range of 0.1-60 cm$^2$V$^{-1}$s$^{-1}$ for several selected device lengths (0.4, 0.6, 1, 2, and 3 μm). The largest enhancement in PCE is seen when the free carrier mobility is increased to 100 cm$^2$V$^{-1}$s$^{-1}$ and the radiative lifetime to 6 ns, values that can be achieved in MoS$_2$ without additional treatments or dielectric engineering.

To understand the impact of electron mobility, we also analyzed the effect of device length. In Figure 3a and Figure S3a,b, we see that the peak PCE occurs at a device length around 1 μm. A similar result is also seen in Figure 3c and Figure S3c,d, where the exciton total lifetime is varied from 0.01 to 6 ns. As these simulations are based on a constant exciton diffusion length of 1.5 μm (other parameters are in Table S4), the optimized device length is comparable to the exciton diffusion length. For lengths greater than 1 μm, the portion of excitons in the i-region diffusing to the dissociation region decreases, leading to a decrease in PCE. For lengths shorter than 1 μm, the exciton diffusion length of 1.5 μm exceeds the dissociation region, also leading to a decrease in PCE. In Figure 3c, we see that increasing the exciton total lifetime significantly increases photocurrent and PCE. In addition to carrier mobility, we also studied the carrier lifetime over the range of 0.5 to 10 ns, where the dominant recombination process at low carrier concentrations is Shockley-Read-Hall (SRH) recombination[48]. In Figure 3d, we see that the PCE increases nonlinearly with carrier lifetime. For SRH lifetimes shorter than 3 ns, the significant increase in PCE shows that the carrier lifetime is a limiting factor for final efficiency. However, for SRH lifetimes longer than 3 ns, the slow increase in PCE indicates that the SRH lifetime is no longer a limiting factor.

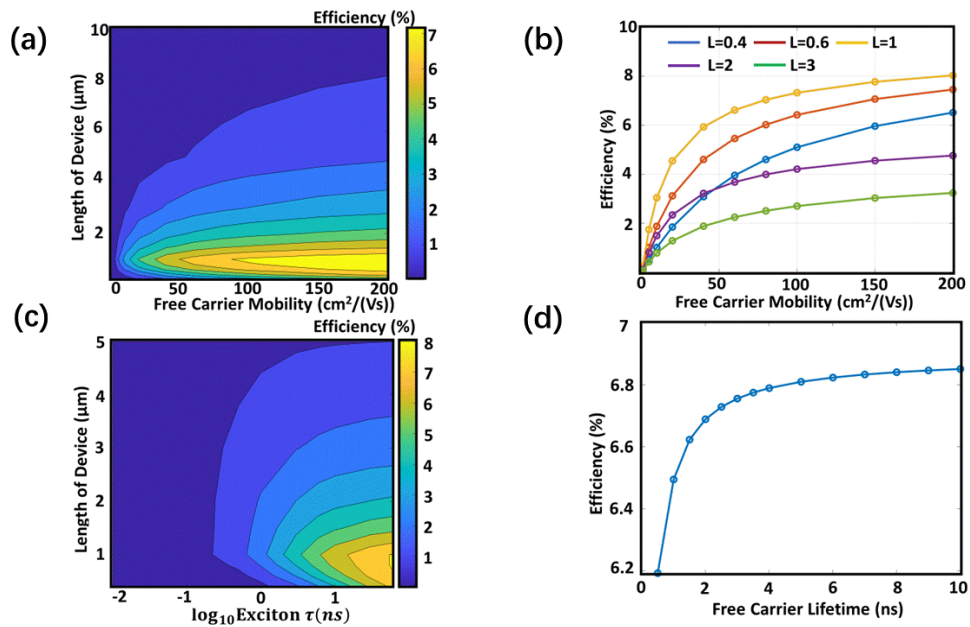

Figure 3. (a) The PCE for the variation of length of device (0.4-10 μm) and free carrier mobility (0.1-200 cm$^2$/(Vs) under exciton lifetime of 10 ns). (b) The PCE for the variation of free carrier mobility for length of device = 0.4 μm, 0.6 μm, 1 μm, 2 μm, 3 μm. (c) The PCE for the variation of length of device (0.2-10 μm) and exciton lifetime (0.01-6 ns) under free carrier mobility of 60 cm$^2$/(Vs)). (d) The PCE for the variation of free carrier lifetime under free carrier mobility of 60 cm$^2$/(Vs).

**Angle sensitivity and comparison between different TMDCs**

The photocarrier generation of the superlattice for N = 1 to 10 with optimized bottom alumina thicknesses was investigated at different incident angles. The photocarrier generation rate was found to have a maximum in the range of 50° to 65° for all values of N. This increased rate is due to the formation of exciton-polaritons, which allows for near unity absorption of transverse electric polarized light near the photon density maximum of the AM1.5 solar spectrum. However, the absorption at other wavelengths decreases as the incident angle increases due to the initial interface becoming more reflective. The maximum occurs at lower incident angles as N increases because the increased Rabi splitting allows for the formation of exciton-polaritons at lower incident angles. Simulations of the absorbed photons at different incident angles were used to calculate the PCE as shown in Figure 4b. The peak PCE of 0.24 eV binding energy was 9.22% at an incident angle of 50°. After optimization, the internal quantum efficiency (IQE) reached a maximum of 69.5%. The IQE did not change when the incident angle was varied in the simulation using the angle dependent photon absorption spectra, which explained the same curvature seen in Figure 4a (N=5) and 4b. The final PCE is proportional to absorbed photons for a constant IQE.

The performance of superlattices made from four different TMDCs, $MoS_2$, $MoSe_2$, $WS_2$, and $WSe_2$, was also simulated using our model. The absorbed photon density was calculated using TMM simulations, as shown in Table 1. Figure 4c shows the absorbed photon density and the PCEs of different material. $MoS_2$ had the largest photon absorption density and the highest PCE due to its two lowest energy excitons being near the maximum of the solar photon spectrum and its relatively low binding energy of 0.24 eV, which resulted in an IQE of 56.72%. $MoSe_2$ had a higher binding energy of 0.57 eV, leading to a significantly lower IQE of 3.79% and a PCE of 0.06%. $WS_2$ had a relatively high free carrier mobility of 1000 $cm^2V^{-1}s^{-1}$ and a low binding energy of 0.32 eV, resulting in the highest IQE of 81.24%. However, its relatively lower photon absorption rate led to a final PCE of 1.98%. $WSe_2$ had a relatively low photon absorption rate and a high binding energy of 0.37 eV, resulting in a PCE of 0.36%. These results show that reducing the exciton binding energy is crucial for producing high-efficiency solar cells using this geometry. The actual binding energy of excitons in the superlattice is expected to be lower by a factor of approximately 2 due to the use of binding energy values for freestanding samples. The binding energy can be further reduced by using high refractive index spacer layers such as hBN, $TiO_2$, and $Ta_2O_5$. Figure 4d shows the current-voltage (I-V) curves of the four materials. In this comparison, $MoS_2$ had the best short-circuit current density of the four TMDCs studied, while $WS_2$ had the largest open-circuit voltage and fill factor (FF).

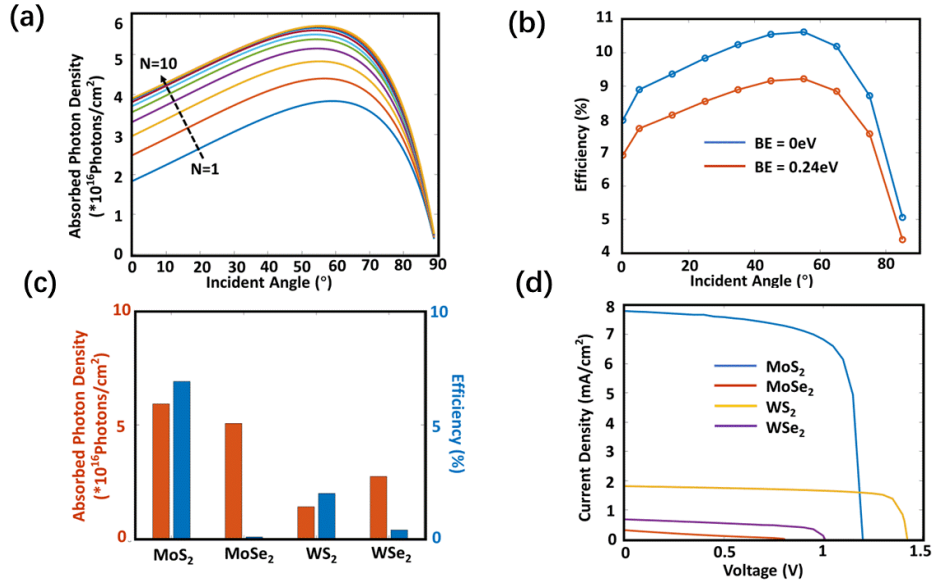

Figure 4. (a) The absorbed photon density for the different incident angle (0-90°) and different layers (N=1-10). (b) The PCE for the variation of the incident angle (0-90°) for binding energy = 0eV, 0.24eV. (c) The absorbed photon density and PCE for $MoS_2$, $MoSe_2$, $WS_2$, and $WSe_2$. (d) The I-V curve for $MoS_2$, $MoSe_2$, $WS_2$, and $WSe_2$.

Table 1. The key parameters of TMDC materials ($MoS_2$, $MoSe_2$, $WS_2$, $WSe_2$).

|  | $MoS_2$ | $MoSe_2$ | $WS_2$ | $WSe_2$ |
|---|---|---|---|---|
| **Absorbed photons (*$10^{16}$ photons/cm$^2$)** | 5.94404 | 5.07596 | 1.39183 | 2.72686 |
| **Bandgap (eV)** | 1.8[18] | 1.60[49] | 2.04[50] | 1.65[49] |
| **Binding Energy (eV)** | 0.24[38] | 0.57[51] | 0.32[30] | 0.37[49] |
| **Mobility (cm$^2$/V/s)** | 60[52] | 480[53] | 1060[54] | 250[55] |
| **Short Circuit Current (mA/cm$^2$)** | 5.40 | 0.31 | 1.81 | 0.68 |
| **Open Circuit Voltage (V)** | 1.20 | 0.82 | 1.42 | 1.01 |
| **Fill Factor** | 70.15 | 22.36 | 76.58 | 53.21 |
| **Internal Quantum Efficiency (%)** | 56.72 | 3.79 | 81.24 | 15.54 |
| **Power Conversion Efficiency (%)** | 4.54 | 0.06 | 1.98 | 0.36 |

At a PCE of 9.22% and a specific weight of 0.58 g m$^{-2}$, excluding a supporting substrate, our device has a specific power (power/weight ratio) of 157 W g$^{-1}$, which is the highest value among TMDC-based cells to the best of our knowledge. We compared the PCE performance and power/weight ratio of our modelled device to other TMDC containing photovoltaic devices from recent literature under AM1.5 illumination (Figure 5)[12-14, 56-70]. Further details on the calculation of specific power for each cell are provided in the Methods section. We found that our solar cell had the largest specific power of all TMDC-based solar cells to the best of our knowledge. This is because previous TMDC-based solar cells have either used monolayer samples for their ideal electronic properties at the expense of lower absorption, or they have used thin films, typically >100 nm, which has large absorption while increasing the weight. However, our design achieves large absorption while also having the advantage of the lightweight, high electrical quality of monolayer TMDCs.

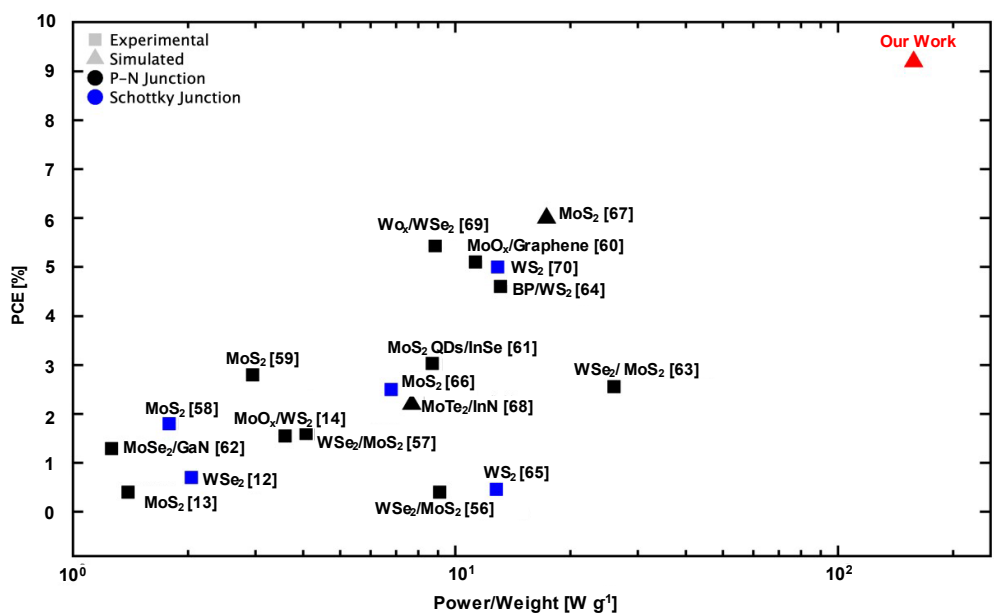

Figure 5. Efficiency chart of vdW material-based photovoltaic with their power/weight ratio in unit of W g$^{-1}$.

We also compared our solar cell to the highest specific power that has been achieved in other materials[71-75] (Figure 6). Although our solar cell has the lowest PCE of all the materials at 9.2% where the rest of the materials range from 11.2% to 22.35%, our exceptionally small active layer thickness of 3.5 nm results in TMDCs having the highest potential specific powers of all of the materials. Since our TMDC-based solar cell has the largest PCE, along with its simplicity and ease of large-scale fabrication, it can be a great candidate for lightweight solar cells in fields such as space, aerospace as well as wearable electronics and remote sensors.

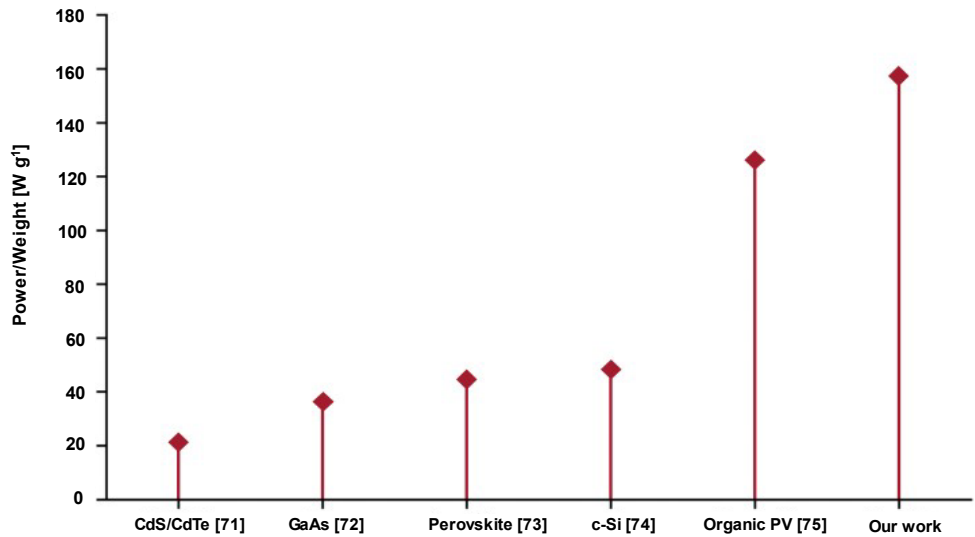

Figure 6. Power/weight ratio chart of different lightweight PV technologies in unit of W g$^{-1}$.

## Conclusion

In summary, we have proposed a device structure and model for 2D TMDC based excitonic solar cells and provided a thorough investigation of the physical factors limiting their performance. The principal innovation of our work lies in the optimization of both the optical and electronic properties and accounting for excitonic effects to estimate the highest possible PCE values using practical materials and device parameters. Our findings suggest that the large exciton binding energies limit the overall efficiencies of 2D TMDC based PV devices to about a third of those predicted by the detailed balance model. None the less, even with large exciton binding energies upon optimizing various geometric and materials quality parameters , a PCE of 9.22% was achieved for MoS$_2$. Even with these PCE values, the total active layer thickness of our proposed optimized device structures is < 4 nm making them some of the highest specific power cells (> 100 W g$^{-1}$) of any thin-film PV technology available today. Overall, our work lays a firm theoretical foundation and computational model on the performance limits of 2D TMDC based excitonic solar cells.

# Methods

**Photocarrier Generation Calculations**

The transfer matrix method[76] was implemented in python to simulate the photocarrier generation rate of the superlattice due to its ability to accurately model the absorption spectra of 1D systems. The refractive index of monolayer TMDCs, $Al_2O_3$, and Au were all taken from literature[77, 78]. The photocarrier generation rate was then calculated by multiplying the absorption spectrum of the superlattice by the AM1.5 solar spectrum[79]. For non-normal incidence, the absorption spectra for TE and TM light were averaged to calculate the absorption spectrum of unpolarized light.

**Sentaurus Simulation**

The two-dimensional solid-state p-i-n superlattice structure was numerically simulated using Sentaurus. This tool is beneficial to elucidate the solar cell behavior considering many physical mechanisms in the device (such as different recombination, different photon absorption calculation). Due to the limitation of vertices number and the long simulation time of the 3D model, a 2D simulation was adopted after we converged on parameters from the 3D model. The 2D simulation has a default thickness of 1 μm in the third dimension where the variation of thickness will not affect the result, details can be found in S3. The structure consists of the monolayer active materials ($MoS_2$, $MoSe_2$, $WS_2$, $WSe_2$), insulators ($Al_2O_3$), cathode (Ag), and anode (Au). The think bottom insulator and gold bottom were not included in the Sentaurus simulation as their effects were fully considered in the photon generation simulation. The investigation of the impact of binding energy, exciton diffusion length, exciton radiative lifetime, exciton nonradiative lifetime, free carrier mobility, free carrier SRH lifetime, and device length on $V_{OC}$, $J_{SC}$, FF and PCE have been performed by utilizing this model.

The initial input parameters were obtained from the experimental data and other theoretical results to define the structure and materials as shown in Table. S1-6. In this model, the electron and hole densities were computed from the electron and hole quasi-Fermi potentials. The bandgap was based on reported values and the Bandgap narrowing effect was not considered. This model allowed discontinuous interfaces for a superlattice structure. By defining the heterointerface, the datasets of two materials were treated properly by introducing double points. The optical generation was based on outer TMM results and was defined manually for each layer. The carrier recombination considered three forms, SRH, auger and radiative. Sentaurus solved the Poisson and continuity equations to account for optical properties. The equation modeled the dynamic of the generation, diffusion, recombination, and radiative decay of singlet excitons[80, 81], is given by:

$$\frac{\partial n_{se}}{\partial t} = R_{bimolec} + \nabla * D_{se}\nabla n_{se} - \frac{n_{se}-n_{se}^{eq}}{\tau} - \frac{n_{se}-n_{se}^{eq}}{\tau_{trap}} - R_{se} \qquad \text{Equ.1}$$

Where $n_{se}$ is the singlet exciton density, $R_{bimolec}$ is the carrier bimolecular recombination rate acting as a singlet exciton generation term, $D_{se}$ is the singlet exciton diffusion constant, $\tau$, $\tau_{trap}$ are the singlet exciton lifetimes. $R_{se}$ is the net singlet exciton recombination rate.

Specific power estimations,

In determining the power/weight ratio for the devices compared in Figure 5 and Figure 6, the specific power (W m$^{-2}$) and specific weight (g m$^{-2}$) are individually calculated. All devices evaluated presented power conversion efficiencies with the AM1.5 spectrum (integrated power = 1000 W m$^{-2}$), thus yielding a specific power that is the PCE fraction of the integrated spectrum power. As there lacks a unified way of calculating specific weight in TMDC photovoltaics, this work views the minimum specific weight sufficient to achieve the given PCE as the comparing metric, positioning this approach specifically as a method for comparing thin film solar cells from a practical optimization perspective. Every layer of a device that actively contributes to power conversion is included in the specific weight, including electrodes which are included based on areal coverage. For layers that do not directly contribute to power conversion but play a role in the optics of the device, such as dielectrics, metal reflectors, and oxide-coated substrates, an effective thickness is used, when smaller than the actual layer thickness, equal to light's penetration depth into the material at the peak of the AM1.5 spectrum (500 nm). Substrates that do not play a role in either the electronic or optical behavior of the device, such as polymer films added for device flexibility, are not included in specific weight calculations given that thicknesses vary across devices without uniform logic. The specific weight is found by multiplying the effective thickness of each included layer in the device by the material density, which was found in literature. Note that the full substrate thickness is used in specific device configurations, such as p-n junction cells where the substrate serves as the p/n side. Dividing specific power by specific weight thus yields the power/weight ratio (W g$^{-1}$).

**Acknowledgements:** D.J., Z. H. and J.L. acknowledge primary support for this work by the Asian Office of Aerospace Research and Development (AOARD) and the Air Force Office of Scientific Research (AFOSR) FA2386-20-1-4074 and FA2386-21-1-4063. D.J. also acknowledges partial support from the Office of Naval Research (N00014-23-1-203) University Research Foundation at Penn and the Alfred P. Sloan Foundation for the Sloan Fellowship. D. L., K. X. and D. J. acknowledges support from the Center for Undergraduate Research Fellowships (CURF) at U. Penn. All authors thank Francisco Barrera for his assistance in setting up simulations and certain material parameters.

# Supporting Information

# How Good Can 2D Excitonic Solar Cells Be?

Zekun Hu,[1] Da Lin,[2] Jason Lynch,[1] Kevin Xu,[1] Deep Jariwala[1,*]

[1]Department of Electrical and Systems Engineering, University of Pennsylvania, Philadelphia, PA, 19104, USA

[2]Department of Materials Science and Engineering, University of Pennsylvania, Philadelphia, PA, 19104, USA

*Corresponding Author (D.J.) email: dmj@seas.upenn.edu

Section 1. Supplementary figures

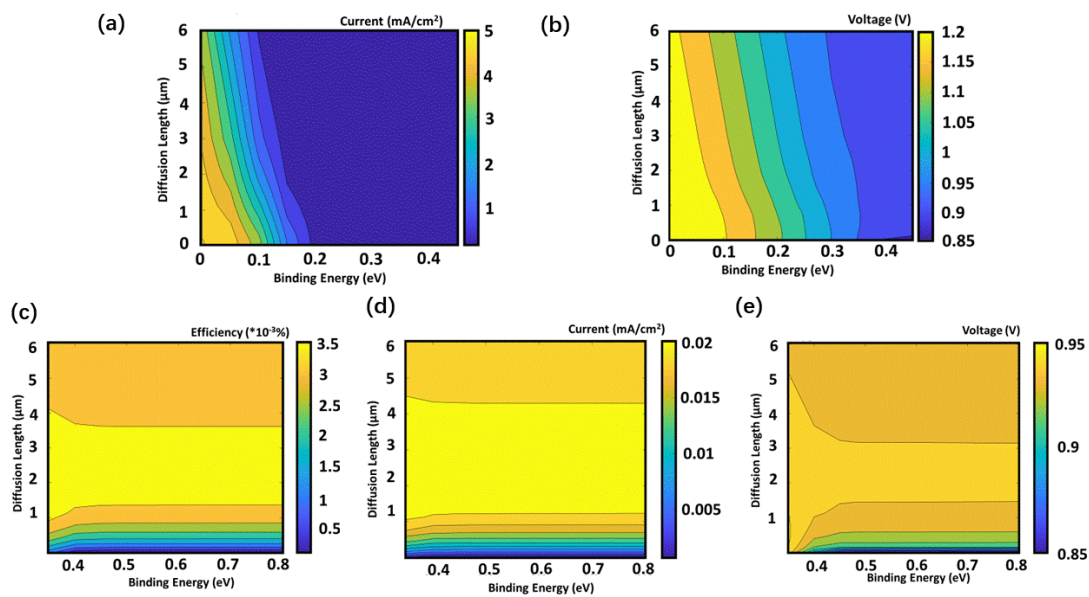

Figure S1. (a) The short-circuit current for the variation of exciton binding energy (0-0.5eV) and diffusion length (0.015-6μm). (b) The open-circuit voltage for the variation of exciton binding energy (0-0.5eV) and diffusion length (0.015-6μm). (c) The PCE for the variation of exciton binding energy (0.3-0.8eV) and diffusion length (0.015-6μm). (d) The short-circuit current for the variation of exciton binding energy (0.3-0.8eV) and diffusion length (0.015-6μm). (e) The open-circuit voltage for the variation of exciton binding energy (0.3-0.8eV) and diffusion length (0.015-6μm). (f) The open-circuit voltage for the variation of exciton radiative lifetime (0.1ps-50ns) and exciton non-radiative lifetime (0.1ps-50ns).

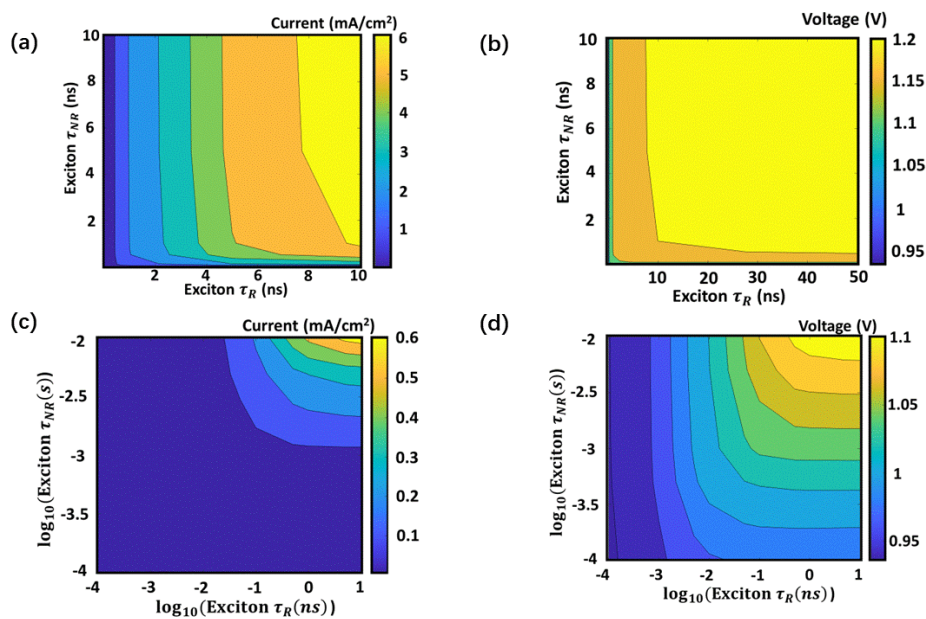

Figure S2. (a) The short-circuit current for the variation of exciton radiative lifetime

(0.1ps-10ns) and exciton non-radiative lifetime (0.1ps-10ns). (b) The open-circuit voltage for the variation of exciton radiative lifetime (0.1ps-50ns) and exciton non-radiative lifetime (0.1ps-10ns). (c) The short-circuit current for the variation of exciton radiative lifetime (0.1ps-10ns) and exciton non-radiative lifetime (0.1ps-10ps). (d) The open-circuit voltage for the variation of exciton radiative lifetime (0.1ps-10ns) and exciton non-radiative lifetime (0.1ps-10ps).

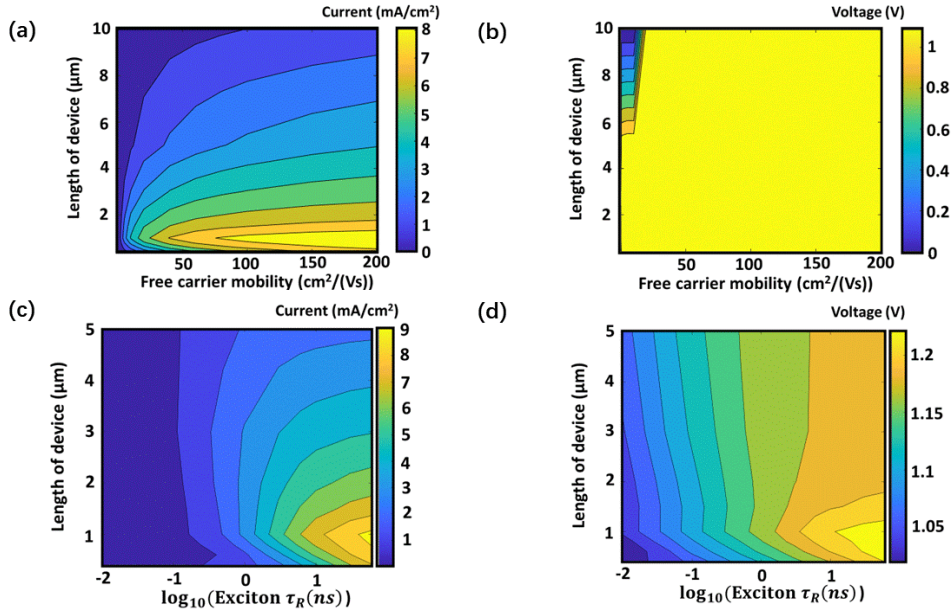

Figure S3. (a) The short-circuit current for the variation of length of device (0.4-10 μm) and free carrier mobility (0.1-200 cm$^2$/(Vs) under exciton lifetime of 10 ns). (b) The open-circuit voltage for the variation of length of device (0.4-10 μm) and free carrier mobility (0.1-200 cm$^2$/(Vs) under exciton lifetime of 10 ns). (c) The short-circuit current for the variation of length of device (0.2-10 μm) and exciton lifetime (0.01-6 ns) under free carrier mobility of 60 cm$^2$/(Vs)). (d) The open-circuit voltage for the variation of length of device (0.2-10 μm) and exciton lifetime (0.01-6 ns) under free carrier mobility of 60 cm$^2$/(Vs)).

Section 2. Supplementary tables

Table S1. Detailed Model of simulations

| Default MoS$_2$ active layer | value |
|---|---|
| Accepter Concentration | 1E-19 cm$^{-3}$ |
| Donor Concentration | 1E-19 cm$^{-3}$ |
| Effective Intrinsic Density | 1.3E4 cm$^{-3}$ |
| Electron Affinity | 4.723 eV |
| Temperature | 300K |
| Radiative Recombination (highest) | 1.841 cm$^{-3}$s$^{-1}$ |
| Auger Recombination (highest) | 7.489E24 cm$^{-3}$s$^{-1}$ |
| SRH Recombination (highest) | 6.986E22 cm$^{-3}$s$^{-1}$ |
| Total Recombination (highest) | 7.674E24 cm$^{-3}$s$^{-1}$ |
| Electron Lifetime | 1.5E-9 s |
| Hole Lifetime | 1.5E-9 s |

Table S2. Tables of parameters in simulations of Figure 2a-b & Figure S1a-e

| | value |
|---|---|
| Material | MoS$_2$ |
| Bandgap (eV) | 1.8[1] |
| Binding energy (eV) | 0-0.8 |
| Diffusion length (microns) | 0.015-6 |
| Exciton_tau_rad (ns) | 0.022[2, 3] |
| Excitom_Tau_nonrad (ns) | 0.002[3, 4] |
| Free carrier SRH lifetime (s) | 1.50E-09[5] |
| Free carrier mobility(cm2/Vs) | 60[6] |
| Device length (microns) | 1 |

Table S3. Tables of parameters in simulations of Figure 2c-d & Figure S2a-d

| | value |
|---|---|
| Material | MoS$_2$ |
| Bandgap (eV) | 1.8[1] |
| Binding energy (eV) | 0.24 |
| Diffusion length (microns) | 1.5 |
| Exciton_tau_rad (ns) | 0.0001-10 |
| Excitom_Tau_nonrad (ns) | 0.0001-10 |
| Free carrier SRH lifetime (s) | 1.50E-09[5] |
| Free carrier mobility(cm2/Vs) | 60[6] |
| Device length (microns) | 1 |

Table S4. Tables of parameters in simulations of Figure 3a-b & Figure S3a-b

| | value |
|---|---|
| Material | MoS$_2$ |
| Bandgap (eV) | 1.8[1] |
| Binding energy (eV) | 0.24 |
| Diffusion length (microns) | 1.5 |
| Exciton_tau_rad (ns) | 0.0001-10 |
| Excitom_Tau_nonrad (ns) | 0.0001-10 |
| Free carrier SRH lifetime (s) | 1.50E-09[5] |
| Free carrier mobility(cm2/Vs) | 60[6] |
| Device length (microns) | 1 |

Table S5. Tables of parameters in simulations of Figure 3c & Figure S3c-d

| | value |
|---|---|
| Material | MoS$_2$ |
| Bandgap (eV) | 1.8[1] |
| Binding energy (eV) | 0.24 |
| Diffusion length (microns) | 1.5 |
| Exciton_tau_rad (ns) | 0.02-12 |
| Excitom_Tau_nonrad (ns) | 0.02-12 |
| Free carrier SRH lifetime (s) | 1.50E-09[5] |
| Free carrier mobility(cm2/Vs) | 60[6] |
| Device length (microns) | 0.4-5 |

Table S6. Tables of parameters in simulations of Figure 3d

| | value |
|---|---|
| Material | MoS$_2$ |
| Bandgap (eV) | 1.8[1] |
| Binding energy (eV) | 0.24 |
| Diffusion length (microns) | 1.5 |
| Exciton_tau_rad (ns) | 20 |
| Excitom_Tau_nonrad (ns) | 20 |
| Free carrier SRH lifetime (s) | 1.50E-09[5] |
| Free carrier mobility(cm2/Vs) | 60[6] |

| Device length (microns) | 1 |

Table S7 Tables of detailed input of different materials

| Material | $MoS_2$ | $MoSe_2$ | $WS_2$ | $WSe_2$ |
|---|---|---|---|---|
| Bandgap (eV) | 1.8[1] | 1.6[7] | 2.04[8] | 1.65[7] |
| Binding Energy (eV) | 0.24[9] | 0.57[10] | 0.32[11] | 0.37[7] |
| Excitons Diffusion length (microns) | 1.5[12] | 0.4[13] | 0.35[14] | 0.16[15] |
| Exciton_tau_rad (ns) | 8[12] | 0.8[16] | 4.4[16] | 3.5[16] |
| Free carrier mobility(cm2/Vs) | 60[6] | 480[13] | 1060[17] | 250[15] |
| Free Carrier Lifetime (ns) | 10[18] | 130[13] | 22[14] | 18[15] |

Section 3. Detailed 3D to 2D convergence simulation.

The simulation was based on a 2D model converged from the 3D model, shown in Figure S4. Initially, a 3D model with larger mesh size was built. Since the third direction duplicates the same structure for times without variation, it has no influence on final results. As shown in Figure S5, different depths tested by the same 3D model show a same PCE. In this case, the 2D simulation with a default depth of 1 micron can correctly represent the 3D model for any depths.

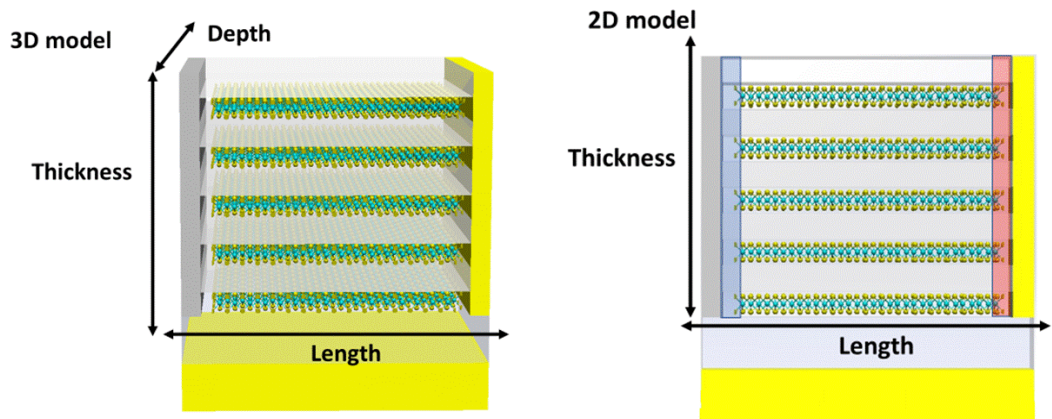

Figure S4. 3D model and 2D model with labelled dimensions.

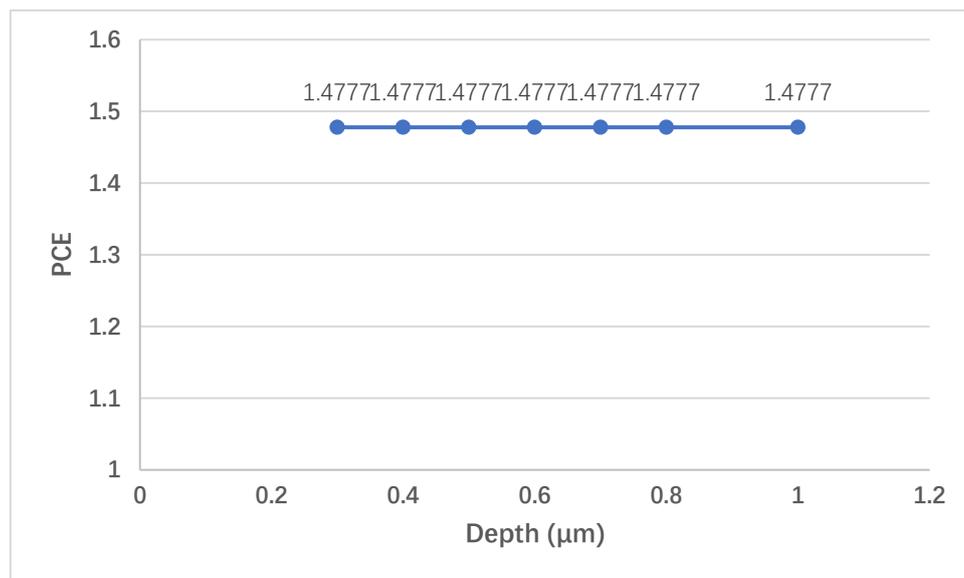

Figure S5. Power conversion efficiency for different depths of the structure in 3D simulations.

The 3D model with a larger mesh size can run a single simulation in an hour. However, as further investigation in Figure S6 shows the results of large mesh size are not reliable. If a smaller mesh size can affect the final results, the current mesh size cannot correctly accurately represent the case. To find a suitable mesh size of the 3D model, a number of simulations was used with smaller mesh sizes. The small mesh size means longer simulation time and heavier computation demand. As we proved the results were not affected by the depth, for some of the small mesh volume, we had to decrease the depth from 1μm to 0.1μm. Still, the simulation of the smallest mesh volume of 2.5E-12 μm$^3$ took more than 24 hours to finish and reached the software tolerance of maximum vertices.

Table S8 Tables of 3D simulation

| Length (μm) | Depth (μm) | Thickness (μm) | Length mesh (μm) | Depth mesh (μm) | Thickness mesh (μm) | mesh volume(μm$^3$) | PCE |
|---|---|---|---|---|---|---|---|
| 1 | 1 | 0.0185 | 0.002 | 0.002 | 0.0005 | 2.00E-09 | 1.28% |
| 1 | 1 | 0.0185 | 0.001 | 0.001 | 0.0005 | 5E-10 | 1.88% |
| 1 | 1 | 0.0185 | 0.0005 | 0.0005 | 0.00025 | 6.25E-11 | 3.79% |
| 1 | 0.1 | 0.0185 | 0.0002 | 0.0002 | 0.00005 | 2E-12 | 6.22% |
| 1 | 0.1 | 0.0185 | 0.001 | 0.0001 | 2.5E-05 | 2.5E-12 | 6.27% |

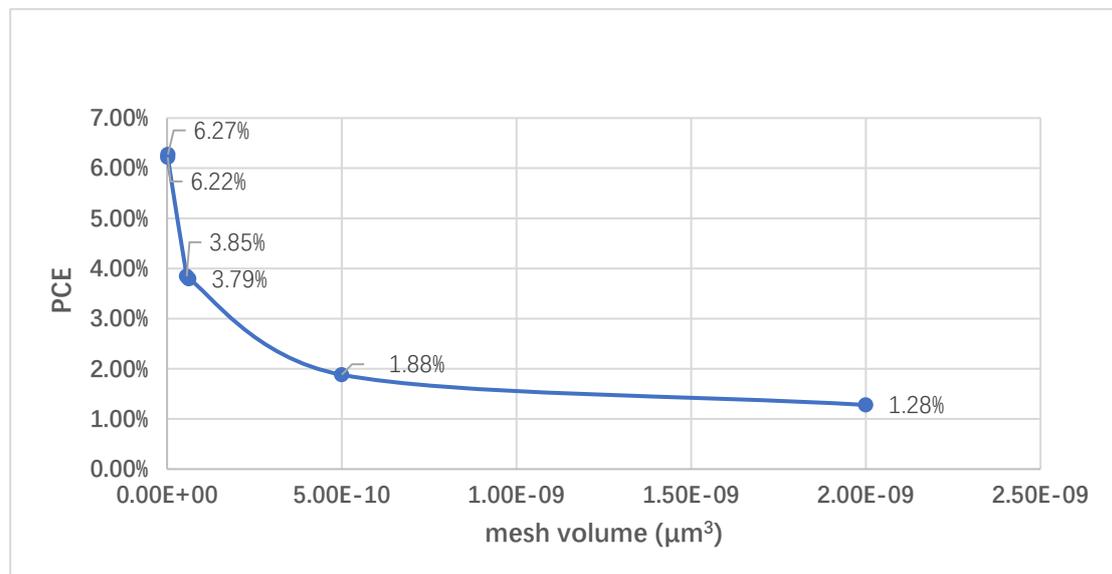

Figure S6. Power conversion efficiency for different mesh volume in 3D simulations.

As the 3D model was not effective to achieve the simulation task, a 2D simulation model was built in Sentaurus. In this software, the default 2D simulation worked as a 3D model with a constant depth of 1 μm, where the depth was not shown on the output. The mesh size of 2D model was optimized to 0.001 μm in length and 0.000025 μm in thickness. As the 0.0185 μm of active layers and insulators was small comparing to the length, a refined mesh size was applied in the direction of thickness. When comparing the 3D simulation and the 2D simulation under the same condition, the 3D simulation with relatively larger mesh lose PCE of 0.2% due to its mesh in the direction of length, shown in Table S9. Even though we did not have exactly same results for both 2D and 3D simulations, the 2D simulation model converged from 3D simulation was valid.

Table S9 Comparison of 2D simulation and 3D simulation with smallest mesh size

|  | Length mesh (μm) | Depth mesh (μm) | Thickness mesh (μm) | PCE |
|---|---|---|---|---|
| 3D simulation | 0.01 | 0.001 | 0.000025 | 6.38% |
| 2D simulation | 0.001 |  | 0.000025 | 6.58% |